\documentclass[aps,prl,showpacs,twocolumn]{revtex4-1}
\usepackage{amssymb}
\usepackage{graphicx}
\usepackage{color}

\begin{document}
\title{Delocalization of Weakly Interacting Bosons in a
1D Quasiperiodic Potential}

\author{V.P. Michal$^{1}$, B.L. Altshuler$^{2}$, and G.V. Shlyapnikov$^{1,3,4}$}
\affiliation{\mbox{$^{1}$ Laboratoire de Physique Th{\'e}orique et Mod{\`e}les Statistiques,  Universit{\' e} Paris Sud, CNRS, 91405 Orsay, France}\\
\mbox{$^{2}$Physics Department, Columbia University, 538 West 120th Street, New York, New York 10027, USA}\\
\mbox{$^{3}$Van der Waals-Zeeman Institute, University of Amsterdam,
Science Park 904, 1098 XH Amsterdam, The Netherlands}\\
\mbox{$^{4}$ Russian Quantum Center, Novaya street 100, Skolkovo, Moscow region 143025, Russia}} 

\date{\today}
\begin{abstract}
We consider weakly interacting bosons in a 1D quasiperiodic potential (Aubry-Azbel-Harper model) in the regime where all single-particle states are localized. We show that the interparticle interaction may lead to the many-body delocalization and we obtain the finite-temperature phase diagram. Counterintuitively, in a wide range of parameters the delocalization requires stronger coupling as the temperature increases. This means that the system of bosons can undergo a transition from a fluid to insulator (glass) state under heating.
\end{abstract}
\maketitle


Quantum mechanics of a single particle in a quasiperiodic potential is a standard although quite nontrivial theoretical problem \cite{Harper1955,Azbel,Aubry1980,Sokoloff,Suslov1982,Thouless1982} intensively studied for several decades. It turned out that for one-dimensional (1D) quasiperiodic potentials (superposition of two incommensurate periodic potentials) the eigenfunctions can be either extended or localized, depending on the parameters of the potential.
The phenomenon of quantum localization was observed for 1D quantum gases in both random \cite{Billy2008} and quasiperiodic potentials \cite{Roati2008}. 
The experiments conducted in the regime of negligible interaction between the 
atoms of expanding Bose gas demonstrated good qualitative agreement with the single-particle theories of localization \cite{Anderson1958, Aubry1980}. The same applies to spreading of wavepackets of light in a quasiperiodic photonic lattice \cite{Lahini}. Recently \cite{Tanzi2013,D'Errico}, a Feschbach resonance was used to study bosons with sizable and 
fine-tunable interaction in the quasiperiodic potential. 

Theoretical description of many-body effects in disordered fermionic \cite{Altshuler1997,Basko2006} and 
bosonic \cite{Aleiner2010} systems is based on the idea of the 
localization of many-body wave functions in the Hilbert space - many-body 
localization. The quasiperiodic potential represents an intermediate case between periodic and disordered systems. Zero temperature phase diagram for 1D bosons in the quasiperiodic potential has been previously discussed and calculated numerically \cite{Roux2008,Deng,Roscilde}, and the case of an infinite temperature for (nearest-neighbor) interacting spinless fermions has been studied in Ref.~\cite{Iyer2013}. The problem of localized and extended states of two interacting particles in the 1D quasiperiodic potential has been discussed in Refs.~\cite{Flach,Dufour}.

In this Letter we study finite-temperature transport properties 
of interacting bosons in the 1D quasiperiodic potential and predict the 
physical behavior which differs drastically from the many-body localization of 
bosons caused by random potentials. We show that, counterintuitively, in a 
broad temperature range an \emph{increase} in temperature induces a transition from 
\emph{fluid to glass}.


The standard model of a 1D quasiperiodic potential is the Aubry-Azbel-Harper 
(AAH) model \cite{Harper1955,Azbel,Aubry1980} - a 
tight-binding Hamiltonian with hopping amplitude $J$ and periodically modulated on-site energies, at a 
period incommensurate with respect to the primary lattice.
The eigenstate $\psi_j^\alpha$ at energy $\varepsilon_\alpha$ 
is determined by the equation
\begin{equation}     \label{AubryModel}
 J(\psi_{j+1}^\alpha+\psi_{j-1}^\alpha)+V\cos(2\pi\kappa 
j)\psi_j^\alpha=\varepsilon_\alpha\psi_j^\alpha.
\end{equation}
Here $V$ is the modulation amplitude, and $\kappa$ is an irrational number. In 1D random potentials all single-particle states are localized \cite{Gertsenshtein1959,Mott1961}. On the contrary, in the AAH model all states are extended unless the amplitude of the modulation exceeds a critical value $2J$. Then all states are localized, and the localization length in units of the lattice constant is given by \cite{Aubry1980}: 
\begin{equation}
\label{zeta}
\!\!\zeta=\frac{1}{\ln(V/2J)}\Rightarrow\,\zeta\simeq\frac{V}{V-2J}\gg 1 \,\,{\rm for} \,\,V-2J\ll V.
\end{equation}
Below we assume $\zeta\gg1$. Our analytical consideration based on the semiclassical approach \cite{Azbel,Suslov1982,Wilkinson1984,Albert2010} to the single-particle problem, is valid when the period of the modulation is much larger than the period of the lattice, $\kappa\ll 1$. We supplement this analysis by numerical calculations for $\kappa\sim 1$, in particular for $\kappa$ equal to the golden ratio $(\sqrt{5}-1)/2$.

The semiclassical one-particle spectrum is organized according to the continued fraction 
decomposition \cite{Azbel,Suslov1982} of the irrational parameter  
$\kappa=1/(n_1+1/(n_2+\dots))$.
For $n_1,n_2,\dots\gg1$ the spectrum has a hierarchical structure: it consists of $n_1$ narrow first-order bands (FOBs) - clusters of $L/n_1$ energy levels ($L$ is the size of the system).  Each FOB contains $n_2$ second-order bands, so that there are  $\sim n_1n_2$ second-order bands in total, $etc$. All eigenstates are located in the energy interval $-V-2J<\varepsilon<V+2J$. The spacing $\omega$ between FOBs is the frequency of the classical periodic motion \cite{LandauLifshitzQM} in a single potential well of the size $1/\kappa\approx n_1$ \cite{n1kappa}, which includes $n_1$ levels:
\begin{equation}     \label{omega}
 \omega=\frac{2\pi}{\oint dx/v}\simeq\frac{2\pi^2\kappa V}{\ln(64V^2/|\varepsilon^2-V^2/\zeta^2|)},
\end{equation}
where $v=\sqrt{4J^2-[\varepsilon-V\cos(2\pi\kappa x)]^2}$ is the classical velocity 
of a particle. 
The widths of FOBs are determined by the tunneling between neighboring wells \cite{Suslov1982}:
\begin{equation}
 \Gamma_s=\frac{4\omega}{\pi}\exp\Big(-\int dx|p|\Big).
\end{equation}
The integral is taken over the classically forbidden region, and 
$|p|=\textrm{arccosh}[(V\cos(2\pi\kappa x)-\varepsilon_s)/2J]$. The index $s$ labels FOBs centered at the energies $\varepsilon _s$. The action can be approximated as $\int dx|p|\approx |\varepsilon_s|/4 \kappa J$, which yields an exponential dependence of the bandwidth on energy: 
\begin{equation}     \label{Gammas}
 \Gamma_s\approx\frac{32\kappa J}{\pi}\exp(-|\varepsilon_s|/4 \kappa J).
\end{equation}

Can the localization of the bosons be destroyed by the interaction? It is known that the interaction 
can delocalize fermions \cite{Basko2006} and 1D bosons \cite{Aleiner2010} in the case of a random potential. Experiments with interacting bosons in 1D quasiperiodic potentials \cite{Tanzi2013,D'Errico} indicated an interaction-induced localization-delocalization transition. It is also worth noting that experiments in quasiperiodic photonic lattices \cite{Lahini} have found that non-linearity (interactions) increases the width of localized wavepackets of light. Here we consider the AAH model with a weak 
on-site interaction: 
\begin{equation}     \label{HB}
 H_{int}=\frac{U}{2}\sum_j a_j^\dagger a_j^\dagger a_j a_j;\,\,\,\,\,U\ll J,
\end{equation}
with $a_j$ being the bosonic field operators. In order to estimate the critical 
coupling constant $U_c$ corresponding to the many-body 
localization-delocalization transition (MBLDT), we use the method developed in 
\cite{Basko2006,Aleiner2010}, which is similar to the original estimation for 
the single-particle Anderson localization \cite{Anderson1958}. One has to consider the localized one-particle states $|\alpha\rangle$ and analyze how different two-particle states $|\alpha,\beta\rangle$ hybridize due to the interaction. The criterion of MBLDT is 
\begin{equation}\label{P}
P_{\alpha}\sim 1,
\end{equation} 
where $P_{\alpha}$ is the probability that for a given one-particle state 
$|\alpha\rangle$ there exist three other states 
$|\beta\rangle,|\gamma\rangle,|\delta\rangle$, such that the two-particle 
states $|\alpha,\beta\rangle$ and $|\gamma,\delta\rangle$ are in resonance, 
i.e. the matrix element $\langle\gamma,\delta|H_{int}|\alpha,\beta\rangle\equiv 
M_{\alpha\beta}^{\gamma\delta}$ exceeds the energy mismatch 
$\Delta_{\alpha\beta}^{\gamma\delta}\equiv 
|\varepsilon_{\alpha}+\varepsilon_{\beta}-\varepsilon_{\gamma}-\varepsilon_{
\delta}|$ where $\varepsilon_{\alpha}$, $\varepsilon_{\beta}$, $\varepsilon_{\gamma}$ and $\varepsilon_{\delta}$ are one-particle energies. 

For large occupation numbers $N_{\beta}$, $N_{\gamma}$, and 
$N_{\delta}$ of the states $|\beta\rangle$, $|\gamma\rangle$ and $|\delta\rangle$ the fluctuations are small. Selecting a given single-particle state $\alpha$ and taking into account both direct and inverse processes we find \cite{note1}
\begin{equation}     \label{melement}
\!\!M_{\alpha\beta}^{\gamma\delta}=\sqrt{|N_{\beta}(1\!+\!N_{\gamma}
)(1\!+\!N_{\delta})\!-\!N_{\gamma}N_{\delta}(1\!+\!N_{\beta})|}U_{\alpha\beta}^{
\gamma\delta},\!\!
\end{equation}
with
\begin{equation}   \label{U}
U_{\alpha\beta}^{\gamma\delta}=U\sum_j\psi_j^{\delta\ast}\psi_j^{\gamma\ast}
\psi_j^\beta\psi_j^\alpha.
\end{equation}
As discussed in Refs.~\cite{Basko2006,Aleiner2010}, the matrix 
elements of the interaction are small unless the energies 
$\varepsilon_{\alpha},\varepsilon_{\beta},\varepsilon_{\gamma},\varepsilon_{
\delta}$ are almost equal pairwise, e.g. 
$\varepsilon_{\alpha}\approx\varepsilon_{\gamma}$ and 
$\varepsilon_{\beta}\approx\varepsilon_{\delta}$, while $\varepsilon_{\alpha}$ 
and $\varepsilon_{\beta}$ can differ substantially. 
Accordingly, $M_{\alpha\beta}^{\gamma\delta}\approx 
N_{\beta}U_{\alpha\beta}^{\gamma\delta}$. The approximation (\ref{melement}) for the matrix elements  
remains valid for small occupation numbers.

If $\alpha$ and $\gamma$ (as well as 
$\beta$ and $\delta$) are nearest neighbors in energy the 
energy mismatch is
\begin{equation}     \label{mismatch}
\!\!\Delta_{\alpha\beta}^{\gamma\delta}\!=\!\delta_{\alpha}\!+\!\delta_{\beta},
\end{equation}
where 
$\delta_{\alpha}=|\varepsilon_{\alpha}-\varepsilon_{\gamma}|$ is a typical spacing between the states on the length scale $\zeta$ at energy close to $\varepsilon_{\alpha}$. 
We estimate the matrix element \cite{SuppMat}:
\begin{equation}     \label{melement1}
M_{\alpha\beta}^{\gamma\delta}\approx UN_{\beta}/\zeta.
\end{equation}

According to Eqs.~(\ref{mismatch}), and (\ref{melement1}) the 
probability $P_{\alpha\beta}^{\gamma\delta}$ of having 
$M_{\alpha\beta}^{\gamma\delta}\gtrsim\Delta_{\alpha\beta}^{\gamma\delta}$ is
\begin{equation}     \label{Palphabeta}
P_{\alpha\beta}^{\gamma\delta}\approx U N_{\beta}/\zeta(\delta_{\alpha} + \delta_{\beta}).
\end{equation}
The probability $P_{\alpha}$ which enters the 
criterion (\ref{P}) of MBLDT is the sum of $P_{\alpha\beta}^{\gamma\delta}$ 
over all single-particle states 
$|\beta\rangle,\,|\gamma\rangle,\,|\delta\rangle$. Since for given $\alpha$ and 
$\beta$ the number of relevant pairs of states 
$|\gamma\rangle,\,|\delta\rangle$ is of order unity, only the summation over 
$\beta$ is important:
\begin{equation}      \label{P1}
\!\!\!P_{\alpha}=\!\!\sum_{\beta,\gamma,\delta}\,\!\! P_{\alpha\beta}^{
\gamma\delta}\!\approx\sum_{\beta}\,\!\!^{\prime}\,UN_{\beta}
/\zeta(\delta_\alpha+\delta_\beta),
\end{equation}
where $\sum\,\!\!^{\prime}$ means that the summation is over the eigenstates on the length scale of $\zeta$.
Substitution of Eq.(\ref{P1}) into Eq.(\ref{P}) leads to the 
criterion of MBLDT:
\begin{equation}     \label{crit}
\sum_{\beta}\,\!\!^{\prime}\, U_cN_{\beta}
/\zeta(\delta_\alpha+\delta_\beta)=1.
\end{equation} 
The critical coupling strength $U_c$ in Eq.(\ref{crit}) depends on the choice of 
the state $\alpha$ through the quantity $\delta_\alpha$. One has to choose $\alpha$ which minimizes $U_c$. Besides, Eq.(\ref{crit}) expresses the critical coupling $U_c$ in terms of the occupation numbers $N_{\beta}$ which are determined by the chemical potential. In order to formulate the MBLDT criterion in terms of the experimentally controllable filling factor $\nu$ one has to complement Eq.(\ref{crit}) with the number equation, which relates $\nu$ to the occupation numbers: 
\begin{equation}   \label{nu}
\nu=\sum_\alpha N_\alpha/L.
\end{equation}

\begin{figure}[h!]
\centering
\includegraphics[width=9cm]{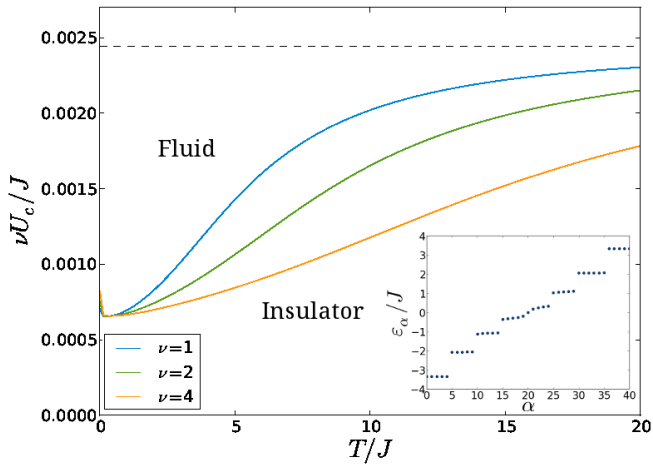}
\caption{The critical coupling strength $U_c$ versus temperature obtained by directly using Eqs.~(\ref{crit})-(\ref{Nalpha}) and the one-particle spectrum computed by exact diagonalization, for $\kappa$ close to $1/8$, $V=2.05J$,  and the filling factor $\nu=1,2,4$. At $T=0$ we recover universality in $\nu U_{c0}$. The dashed line is the $T\to\infty$ asymptotics. The inset shows the spectrum on the length scale of $\zeta$ (the number of states is $\zeta\approx 40$).}
\label{nuUc43d350V205}
\end{figure}

We are now ready to apply the MBLDT criterion to interacting bosons in the AAH model.
On the insulator side, the occupation numbers are given by:
\begin{equation}   \label{Nalpha}
N_\alpha=\{\exp[(\varepsilon_\alpha-\mu+UN_\alpha
/\zeta)/T]-1\}^{-1},
\end{equation}
where $\mu$ is the chemical potential shifted by the interaction energy of a particle in the state $|\alpha\rangle$ with particles in all other states.
Equations (\ref{crit}), (\ref{nu}), and (\ref{Nalpha}) yield the critical coupling $U_c$ as a function of temperature.

On the length scale $\zeta$ there are $\sim \zeta$ states with a significant amplitude of the wavefunction. Thus, there are at most $\zeta$ states contributing to the sum in Eq.(\ref{crit}). The calculation simplifies in the limit $n_1\ll \zeta\ll n_1n_2$ where each FOB contains $\sim\zeta/n_1\approx \kappa\zeta$ overlapping states. The related example is shown in Fig.1: $\kappa$ is close to $1/8$, the localization length is $\zeta\simeq 40$, and $n_2=7$. 

At temperatures much smaller than the spacing between the FOBs, i.e. $T\ll\omega$, only single-particle states from the lowest energy FOB participate in MBLDT. We assume that the spacing between the states in this band is approximately constant and thus equal to $\delta_{\beta}\approx\Gamma_0/\kappa\zeta$. Using the fact that the sum $\sum_{\beta}\,\!\!^{\prime}\,N_{\beta}/\zeta$ over the states on the length scale $\zeta$ in Eq.(\ref{crit}) is equal to the sum $\sum_{\alpha}N_{\alpha}/L$ in Eq.(\ref{nu}) over all states we obtain \cite{infact}:
\begin{equation}    \label{UclowT}
\nu U_{c}\approx 2\Gamma_0/\kappa\zeta;\,\,\,\,\,T\ll\omega.
\end{equation}

At temperatures $T\gg\omega$ many FOBs are occupied and particles in these bands participate in MBLDT. The bands are so narrow that all levels in the $s$-th FOB have the same occupation, so that the corresponding level spacing is $\delta_{\beta}\approx\Gamma_s/\kappa\zeta$. Then, summing over $\beta$ within each FOB one can rewrite Eq.(\ref{crit}) as
\begin{equation}    \label{sumsU}
 \sum_{s=0}^{n_1-1}\zeta U_cN_s\kappa^2/(\Gamma_0+\Gamma_s)\approx 1
\end{equation}
(we selected the state $|\alpha\rangle$ to be in the lowest  energy band, $s=0$).
According to Eq.(\ref{Gammas}) the width $\Gamma_s$ exponentially increases with $\varepsilon_s$ (i.e. $s$) in the interval $0>\varepsilon_s>\varepsilon_0\simeq -V-2J$, since $|\varepsilon_s|$ decreases. Hence, at $T\ll8J$ the sum over $s$ in Eq.(\ref{sumsU}) is dominated by $s=0$, which leads to
\begin{equation}\label{scrit}
U_c\approx2\Gamma_0/N_0\kappa^2\zeta.
\end{equation} 

For $\varepsilon_s-\mu<T$ we may use the occupation number expression \cite{SuppMat}  
\begin{equation}    \label{Nsbig}
N_s\approx T/(\varepsilon_s-\mu),
\end{equation}
and put $N_s\approx0$ for larger $\varepsilon_s$. In particular $N_0\approx T/(\varepsilon_0-\mu)$. With the use of Eq.(\ref{Nsbig}), one can rewrite Eq.(\ref{nu}) as
\begin{equation}
 \nu\approx\sum_{s=0}^{(T+\mu-\varepsilon_0)/\omega}\kappa T/(s\omega+\varepsilon_0-\mu).
\end{equation}
In the temperature range $\omega\!\ll\! T\ll 8J$ we then find the chemical potential dependence on temperature \cite{SuppMat}
\begin{equation} \label{muT}
 \mu\approx\varepsilon_0-\frac{\kappa T}{\nu}\Big[1+\frac{T}{8\nu J}\ln\Big(\frac{T}{\omega}\Big)\Big],
\end{equation}
and using equation (\ref{scrit}) we obtain the critical coupling:
\begin{equation} \label{UT}
 \nu U_c\approx\frac{2\Gamma_0}{\kappa\zeta}\Big[1\!+\!\frac{T}{8\nu J}\!\ln\!\Big(\frac{T}{\omega}\Big)\Big].
\end{equation}
Since $T\ll 8J$, the second term in square brackets is a small correction. Nevertheless, it is important. According to Eq.(\ref{UT}) the temperature dependence of the critical coupling is anomalous: $U_c$ increases with $T$, i.e. an increase in temperature favors the insulator state.

\begin{figure*}
\includegraphics[width=18cm]{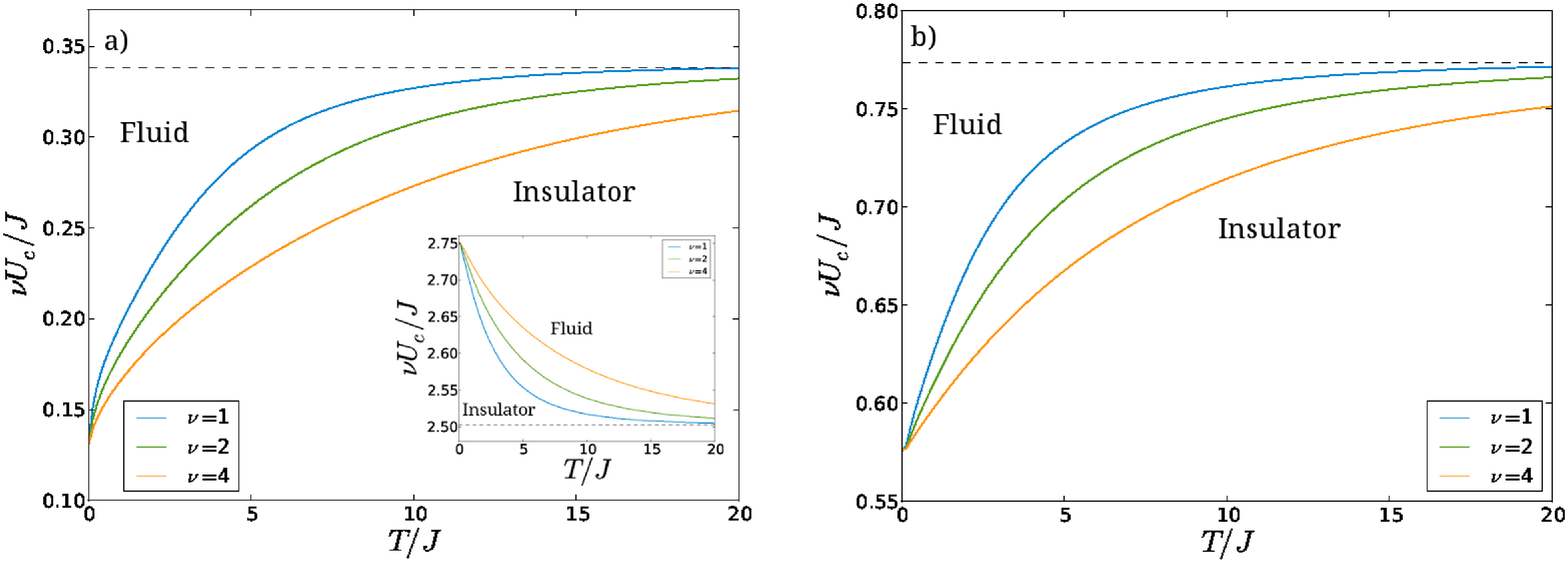}
\caption{The same as in Fig.\ref{nuUc43d350V205} for $V=2.3J$ ($\zeta\approx 7$). In (a) $\kappa\approx 0.24$, and in (b) $\kappa$ is equal to the golden ratio. The inset in (a) shows $\nu U_c(T)$ for $\!\kappa\!\approx\! 1/8$ and $V\!=\!2.25J$ ($\zeta\!\approx\! 8$).}
\label{nuUc233d377V23}
\end{figure*}

This behavior originates from the cluster structure of the spectrum, with exponentially increasing cluster width when going from the lowest (highest) cluster energy to the middle of the spectrum. Therefore, at $T\ll 8J$ (and $\zeta\gg n_1\approx\kappa^{-1}$) the localization-delocalization transition is provided only by the particle states in the lowest energy cluster. The fraction of these particles decreases with increasing temperature, thus ensuring an increase in the critical coupling strength $U_c$.

For temperatures $T\rightarrow\infty$ ($T\gg 8J, 8\nu J$) all eigenstates are equally populated and $N_s=\nu$. Then the main contribution to the sum in Eq.(\ref{sumsU}) comes from $s=0$ and $s=n_1-1$ as $\Gamma_{n_1-1}\approx\Gamma_0\approx(32\kappa J/\pi)\exp(-1/\kappa)$. Having in mind Eq.(\ref{UclowT}) we thus obtain:  
\begin{equation}    \label{Uinfty}
\nu U_{c\infty}\simeq \Gamma_0/\kappa^2\zeta\simeq \nu U_{c0}/2\kappa\gg \nu U_{c0},
\end{equation}
where $U_{c0}$ is the zero temperature critical coupling at $\zeta\gg n_1\approx\kappa^{-1}$. Therefore, under this condition we always expect the anomalous ''freezing with heating'' behavior at high enough temperatures.

Our analytical results are confirmed by numerics using the single-particle spectrum obtained by exact diagonalization. The results for $\kappa\approx 1/8$ and $\zeta\simeq 40$ are shown in Fig.1.
The critical coupling turns out to be very small since it is proportional to the width $\Gamma_0$, which is several orders of magnitude smaller than $J$. This justifies the validity of our perturbative approach with respect to the interparticle interaction.

In the opposite limit $1\ll\zeta\lesssim\kappa^{-1}$, the situation changes. Indeed in this case single-particle states participating in MBLDT belong to different FOBs (not more that one state from a given cluster). The characteristic spacing between these states is $\sim 8J/\zeta$ and the cluster structure of the spectrum is not important. The resulting critical coupling $\nu U_c$ is $\sim J$. For $\kappa\ll 1$ one can use the quasiclassical approach with the density of states $\kappa/\omega(\varepsilon)$ and  $\omega(\varepsilon)$ given by Eq.(\ref{omega}). The results are consistent with our calculations based on exact diagonalization for the one-particle spectrum and Eqs.~(\ref{crit}), (\ref{nu}), and (\ref{Nalpha}). They suggest a slow decrease of $U_c$ with increasing temperature and are displayed in the inset of Fig.2a for $\zeta$ and $\kappa^{-1}$ both close to $8$. However, since our approach is based on the perturbative treatment of the interactions, its predictions at $\nu U_c\gtrsim J$ at least 
require a large filling factor $\nu$. A detailed analysis of this question will be given elsewhere.

For $\kappa\sim 1$ the quasiclassical approach is no longer valid and one has to rely only on the numerics based on exact diagonalization for the one-particle problem and Eqs.~(\ref{crit}), (\ref{nu}), and (\ref{Nalpha}). The results for $\kappa=1/(1+1/(1+\dots))=(\sqrt{5}-1)/2$ (golden ratio) and for $\kappa$ close to $0.24$ at $\zeta\approx 7$ are shown in Fig.2. For the latter case our results at $T=0$ and $\nu=1$ are consistent with the DMRG calculations of Refs.~\cite{Roux2008} and \cite{Deng} (using $\kappa=0.77$ which is equivalent to $\kappa=0.23$) extrapolated to $V=2.3J$. 

Ref. \cite{Iyer2013} presented results of the numerical simulation for spinless fermions with nearest neighbor interaction subject to a quasiperiodic potential at $T=\infty$. This problem (different from bosons with the onsite interaction), can also be attacked with our approach at any temperature. The results will be published elsewhere, but already now we can say that at $T\to\infty$ they agree fairly well with Ref. \cite{Iyer2013}.

Our results at finite temperatures indicate an anomalous $U_c(T)$-dependence. The experiment \cite{Tanzi2013} has been performed for $\kappa\approx 1.24$, which according to Eq.(\ref{AubryModel}) is equivalent to $\kappa=0.24$. The extrapolation of experimental results to $V\approx 2.3J$ gives $\nu U_c/J\sim 0.3$, which is consistent with our calculations. 

In conclusion, we have developed the many-body localization theory of weakly interacting bosons in a 1D quasiperiodic potential and obtained the phase diagram in terms of temperature and interaction. The most unexpected prediction based on our calculations is the transition from fluid to insulator (glass) with heating. 

We are grateful to I.L. Aleiner and G. Modugno for fruitful discussions and acknowledge support from Triangle de la Physique through the project DISQUANT, support from IFRAF and from the Dutch Foundation FOM. The research leading to these results has received funding from the European Research Council under European Community's Seventh Framework Programme (FR7/2007-2013 Grant Agreement no.341197).


\begin{thebibliography}{99}
\bibitem{Harper1955} P.G. Harper, Proceedings of the Physical Society. Section A 
{\bf 68}, 874 (1955).
\bibitem{Azbel} M.Ya. Azbel, Sov. Phys. JETP {\bf 17}, 665 (1963), Sov. Phys. 
JETP {\bf 19}, 634 (1964), Phys. Rev. Lett. {\bf 43}, 1954 (1979).
\bibitem{Aubry1980} S. Aubry and G. Andr\'e, Ann. Isr. Phys. Soc. {\bf 3}, 133 
(1980).
\bibitem{Sokoloff} J.B. Sokoloff, Phys. Rev. B {\bf 22}, 5823 (1980); {\it ibid} {\bf 23}, 6422 (1981); Physics Reports {\bf 126}, 189 (1985). 
\bibitem{Suslov1982} I.M. Suslov, Sov. Phys. JETP {\bf 56}, 612 (1982). In this paper the notation $\Gamma_s$ is used for the halfwidth.
\bibitem{Thouless1982} D.J. Thouless, M. Kohmoto, M.P. Nightingale, and M. den Nijs, Phys. Rev. Lett. {\bf 49}, 405 (1982). 
\bibitem{Billy2008} J. Billy, V. Josse, Z. Zuo, A. Bernard, B. Hambrecht, P. Lugan, D. Clement, L. Sanchez-Palencia, P. Bouyer, and A. Aspect, Nature {\bf 453}, 891 (2008).
\bibitem{Roati2008} G. Roati, C. D'Errico, L. Fallani, M. Fattori, C. Fort, M. Zaccanti, G. Modugno, M. Modugno, and M. Inguscio, Nature {\bf 453}, 895 (2008).
\bibitem{Anderson1958} P.W. Anderson, Phys. Rev. {\bf 109}, 1492 (1958).
\bibitem{Lahini} Y. Lahini, R. Pugatch, F. Rozzi, M. Sorel, R. Morandotti, N. Davidson, and Y. Silberberg, Phys. Rev. Lett. {\bf 103}, 013901 (2009).
\bibitem{Tanzi2013} L. Tanzi, E. Lucioni, S. Chaudhuri, L. Gori, A. Kumar, C. 
D'Errico, M. Inguscio, and G. Modugno, Phys. Rev. Lett. {\bf 111}, 115301 
(2013).
\bibitem{D'Errico} C. D'Errico, E. Lucioni, L. Tanzi, L. Gori, G. Roux, I.P. 
McCulloch, T. Giamarchi, M. Inguscio, and G. Modugno, submitted for publication.
\bibitem{Altshuler1997} B.L. Altshuler, Yu. Gefen, A. Kamenev, and L.S. Levitov, Phys. Rev. Lett. {\bf 78}, 2803 (1997).
\bibitem{Basko2006} D. Basko, I.L. Aleiner, and B.L. Altshuler, Annals of Physics {\bf 321}, 1126 (2006); {\it Problems of Condensed Matter Physics}, Editors A.L. Ivanov and S.G. Tikhodeev, Chapter {\it On the problem of many-body localization}, (Oxford University Press, 2007).
\bibitem{Aleiner2010} I.L. Aleiner, B.L. Altshuler, and G.V. Shlyapnikov, Nature Physics {\bf 6}, 900 (2010).
\bibitem{Roux2008} G. Roux, T. Barthel, I.P. McCulloch, C. Kollath, U. Schollw\"ock, and T. Giamarchi, Phys. Rev. A {\bf 78}, 023628 (2008).
\bibitem{Deng} X. Deng, R. Citro, A. Minguzzi and E. Orignac, Phys. Rev. A {\bf 78}, 013625 (2008).
\bibitem{Roscilde} T. Roscilde, Phys. Rev. A {\bf 77}, 063605 (2008).
\bibitem{Iyer2013} S. Iyer, V. Oganesyan, G. Refael, and D. A. Huse, Phys. Rev. 
B {\bf 87}, 134202 (2013).
\bibitem{Flach} S. Flach, M. Ivanchenko, and R. Khomeriki, European Phys. Lett. {\bf 98}, 66002 (2012).
\bibitem{Dufour} G. Dufour and G. Orso, Phys. Rev. Lett. {\bf 109}, 155306 (2012).
\bibitem{Gertsenshtein1959} M. E. Gertsenshtein and V. B. Vasiliev, Theory Probab. Appl. {\bf 4}, 391 (1959).
\bibitem{Mott1961} N. F. Mott and W. D. Twose, Adv. Phys. {\bf 10}, 107 (1961).
\bibitem{Wilkinson1984} M. Wilkinson, Proc. R. Soc. London Ser. A {\bf 391}, 305 (1984).
\bibitem{Albert2010} M. Albert and P. Leboeuf, Phys. Rev. A {\bf 81}, 013614 (2010).
\bibitem{LandauLifshitzQM} L. D. Landau and E. M. Lifshitz, Quantum Mechanics
(Non-relativistic Theory), Course of Theoretical Physics Volume 3 Third 
Edition, Elsevier (1977).
\bibitem{n1kappa} Hereinafter considering the limit $\kappa\ll 1$ we put everywhere $n_1=\kappa^{-1}$. 
\bibitem{note1} The matrix element $\langle\gamma,\delta|H_{int}|\alpha,\beta\rangle$ by itself is proportional to $\sqrt{N_{\alpha}}$. However, the related probability (proportional to $N_{\alpha}$) determines the time derivative of $N_{\alpha}$, so that $N_{\alpha}$ drops from the expression for the probability of hybridization of a given state of the $N_{\alpha}$-manifold. This is equivalent to considering a single particle state $|\alpha\rangle$ with $N_{\alpha}=1$. 
\bibitem{SuppMat} See Supplementary Materials.
\bibitem{infact} In fact, the level spacing $\delta_{\beta}$ is minimal in the center of the FOB and somewhat increases closer to the band edges. 
This slightly increases the critical coupling compared to the result of Eq.(\ref{UclowT}). However, the quantity $\nu U_c\kappa\zeta/2\Gamma_0$ remains equal to unity within a factor of 2 up to $\kappa\approx 0.25$, where the cluster structure of the spectrum is still pronounced but the quasiclassical approach is strictly speaking not valid. 
A decrease in the level spacing with increasing energy in the low-energy part of the FOB can also influence the temperature dependence of $U_c$ at $T<\omega$. Indeed, as shown in Fig. \ref{nuUc43d350V205}, the curve $U_c(T)$ has a dip at very low temperatures. If $T<\omega$ and the occupation of the lowest energy FOB is large, the critical coupling $U_c$ first decreases with increasing temperature (this is analogous to the behavior of $U_c(T)$ displayed in the inset of Fig.2, where not more than one state of a given FOB participates in MBLDT). Once the temperature exceeds the spacing between the FOBs the exponential dependence of the FOB widths dominates and the temperature dependence of $U_c$ becomes anomalous.

\end{thebibliography}
\end{document}